\documentclass[runningheads]{llncs}
\usepackage[T1]{fontenc}
\usepackage{graphicx,verbatim}
\usepackage{amsmath}
\usepackage{enumitem}
\usepackage{hyperref}
\usepackage{color}

\begin{document}
\title{Pretext Task Adversarial Learning for Unpaired Low-field to Ultra High-field MRI Synthesis}
\titlerunning{PTA Framework for Ultra High-field MRI Synthesis}
\author{Zhenxuan Zhang\inst{1}* \and
Peiyuan Jing\inst{1}\thanks{Equal contribution.} \and
Coraline Beitone\inst{1} \and
Jiahao Huang\inst{1,2} \and
Zhifan Gao\inst{5} \and
Guang Yang\inst{1,2,3,4}** \and
Pete Lally\inst{1}\thanks{Co-corresponding author.}}
\authorrunning{Zhenxuan Zhang et al.}
\institute{Bioengineering Department and Imperial-X, Imperial College London, London W12 7SL, UK \and
National Heart and Lung Institute, Imperial College London, London SW7 2AZ, UK \and
Cardiovascular Research Centre, Royal Brompton Hospital, London SW3 6NP, UK \and
School of Biomedical Engineering \& Imaging Sciences, King's College London, London WC2R 2LS, UK \and
School of Biomedical Engineering, Sun Yat-sen University, Guangzhou 510006, China\\\email{g.yang@imperial.ac.uk, p.lally@imperial.ac.uk}}


\maketitle              
\begin{abstract}
Given the scarcity and cost of high-field MRI, the synthesis of high-field MRI from low-field MRI holds significant potential when there is limited data for training downstream tasks (e.g. segmentation). Low-field MRI often suffers from a reduced signal-to-noise ratio (SNR) and spatial resolution compared to high-field MRI. However, synthesizing high-field MRI data presents challenges. These involve aligning image features across domains while preserving anatomical accuracy and enhancing fine details. To address these challenges, we propose a Pretext Task Adversarial (PTA) learning framework for high-field MRI synthesis from low-field MRI data. The framework comprises three processes: (1) The slice-wise gap perception (SGP) network aligns the slice inconsistencies of low-field and high-field datasets based on contrastive learning. (2) The local structure correction (LSC) network extracts local structures by restoring the locally rotated and masked images. (3) The pretext task-guided adversarial training process introduces additional supervision and incorporates a discriminator to improve image realism. Extensive experiments on low-field to ultra high-field task demonstrate the effectiveness of our method, achieving state-of-the-art performance (16.892 in FID, 1.933 in IS, and 0.324 in MS-SSIM). This enables the generation of high-quality high-field-like MRI data from low-field MRI data to augment training datasets for downstream tasks. The code is available at: \url{https://github.com/Zhenxuan-Zhang/PTA4Unpaired_HF_MRI_SYN}.

\keywords{High-field MRI  \and Image Synthesis \and Self-supervised Method.}

\end{abstract}
\section{Introduction}
Given the scarcity and cost of high-field MRI, the synthesis of high-field MRI has significant potential to augment training datasets for downstream tasks. Low-field MRI (below 1.5T) is commonly used in resource-limited settings due to its lower cost and compact equipment size \cite{unity,low-field}. However, it produces lower-quality images with reduced signal-to-noise ratio and spatial resolution, limiting its effectiveness in detecting subtle or complex pathologies. In contrast, high-field MRI (typically 3T and above) offers superior image quality and spatial resolution \cite{Ultra-high}, allowing for detailed visualization of small lesions and accurate quantitative analysis. However, access to high-field MRI remains limited due to equipment availability and technical expertise requirements \cite{SACE,SynGAN}. Although large high-field MRI datasets at 3T are available for training deep learning models (e.g., fast MRI dataset \cite{fastMRI}), expanding these datasets to cover more general pathologies and higher field strengths remains a challenge \cite{SynGAN,WATNet,SACE}. Therefore, synthesizing high-field MRI data from low-field MRI data holds significant promise, particularly for large-scale unpaired datasets \cite{UNest,Cyclegan}. This approach leverages the accessibility of low-field MRI while overcoming its inherent limitations in image quality, giving an opportunity to augment the limited datasets available at high-field, providing a larger training set for downstream tasks. 
\begin{figure*}[t]
\centerline{\includegraphics[width=\columnwidth]{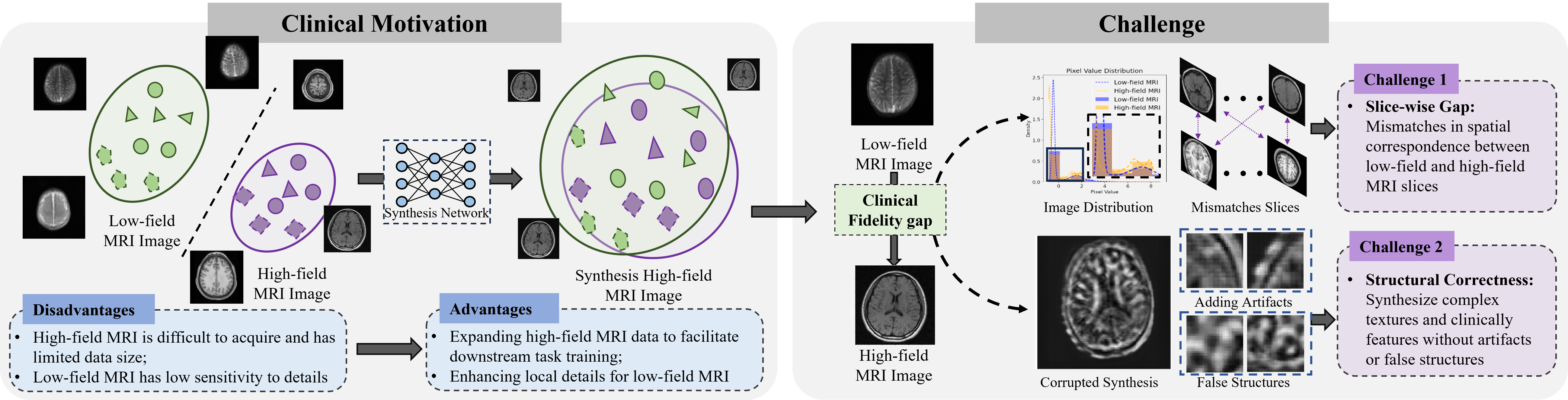}}
\caption{The motivation and challenges of our PTA framework. (a) Clinical motivation lies in synthesizing high-field MRI from low-field MRI. (b) The challenge is the clinical fidelity gap between low-field MRI data and high-field MRI data. }
\label{fig1}
\end{figure*}

However, synthesizing high-field MRI from unpaired low-field data remains challenging due to the clinical fidelity gap \cite{low-field,lisa,m4raw}, which involves slice-wise alignment and structural-detail accuracy. First, the slice-wise gap arises from spatial mismatches in unpaired datasets \cite{resvit,RIRGAN}, where a low-field slice may be misaligned with a high-field slice from a different anatomical position (e.g., a mid-brain slice paired with a top-of-head slice). This misalignment distorts spatial references, degrading synthesis quality. Second, structural-detail accuracy is critical for preserving anatomical fidelity and fine details \cite{UNest,Cyclegan}. Misrepresented structures, such as brain lesions, pose risks if used for diagnosis, especially with noisy low-field inputs \cite{fastMRI,HCP}. High-field MRI contains intricate textures and clinically relevant features (e.g., lesion contrast) that are difficult to synthesize without artifacts due to the ill-posed nature of the problem. However, for downstream tasks (e.g., tissue segmentation), realistic feature synthesis can enrich limited high-field datasets. Thus, an effective model must balance diversity while ensuring clinical realism.

Existing methods struggle to address the clinical fidelity gap in low-to-high-field MRI synthesis. While GAN-based approaches have shown promise in medical image translation \cite{SynGAN,RIRGAN,CFGAN,esrgan,Cyclegan}, they often fail to resolve the slice-wise gap and local structural accuracy. Many methods rely on paired data to mitigate spatial mismatches by learning from aligned slices \cite{SynGAN,RIRGAN,CFGAN,esrgan}. However, in unpaired settings, ensuring anatomical alignment in synthesized high-field images remains challenging \cite{cytran,UNest}. Without explicit slice-wise correspondence, models risk generating misaligned or anatomically unrealistic details (e.g., incorrect lesion positioning or structural distortions) \cite{Cyclegan}.
Additionally, many GAN-based models prioritize visual realism over clinical relevance \cite{Cyclegan,cytran,hyper-GAN,UNest}. While images may appear plausible, they often lack crucial diagnostic details, particularly in complex regions like brain lesions and small anatomical structures. This compromises anatomical fidelity, leading to misrepresented or lost clinically significant features. The key challenge remains to generate high-field MRI data that are not only visually convincing but also clinically reliable.

\begin{figure*}[t]
\centerline{\includegraphics[width=\columnwidth]{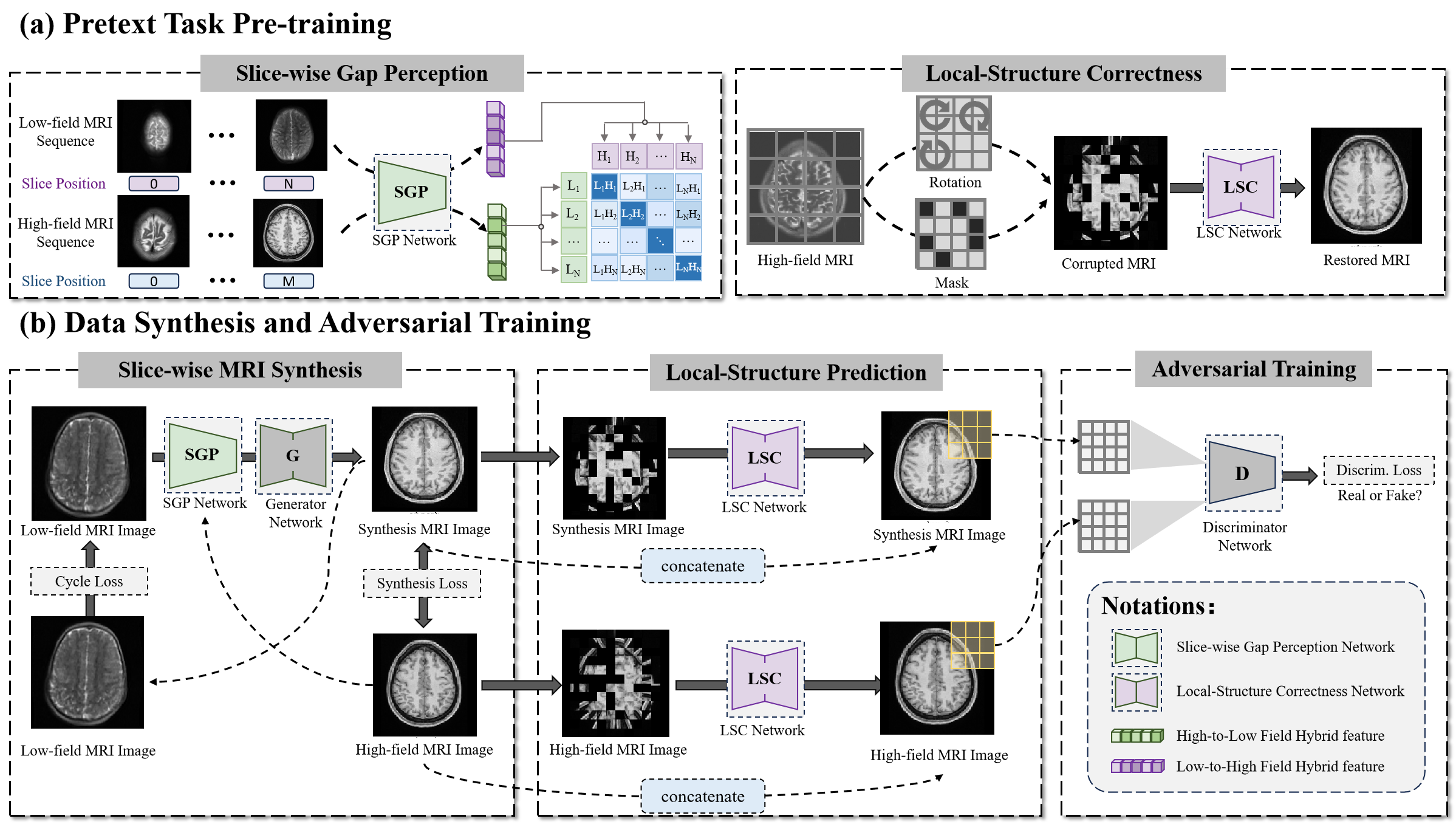}}
\caption{Workflow of the proposed  pretext task adversarial (PTA) learning framework. (a) Pretext Task Pre-training focuses on reducing domain gaps between low-field and high-field MRI through slice-wise gap perception and local-structure correctness. (b) Data Synthesis and Adversarial Training aims to synthesize high-field MRI data from low-field MRI inputs using a synthesis network. }
\label{fig2}
\end{figure*}
In this paper, we propose a pretext task adversarial (PTA) learning framework for high-field MRI synthesis (Figure\ref{fig2}) . Unlike previous approaches that rely on paired low-to-high-field MRI data (specific sequence order and details), our method introduces slice-level alignment and local detail correctness constraints, allowing for better generalization in unpaired synthesis. Slice-wise gap perception leverages sequence contrast learning to analyze consecutive low-field and high-field MRI slices. This ensures that the most relevant high-field images are selected as references during synthesis, mitigating spatial mismatches. Local structure correction employs a self-supervised learning approach, where details are recovered from locally rotated and masked images. This serves as a pretext task, providing additional local supervision to enhance fine structural accuracy. Notably, when errors or artifacts distort local details, the pretext task fails to recover them, making it easier for the discriminator in the adversarial training process to assess image correctness. 
These modules establish a robust framework for improving the usability of low-field MRI data in downstream applications which require high-field MRI data. Our contributions lie in three folds:
\begin{enumerate}
    \item To the best of our knowledge, our work is the first unpaired LF-to-HF MRI synthesis framework to eliminate the reliance on scarce paired data.
    \item We propose a unified framework that integrates slice-wise gap perception, local structure correction, and pretext task-guided adversarial training to enhance high-field MRI synthesis.
    \item Extensive experiments on low-field to ultra high-field task demonstrate the state-of-the-art performance of PTA, achieving 16.892 in FID, 1.933 in IS, and 0.324 in MS-SSIM.
\end{enumerate}

\begin{figure*}[t]
\centerline{\includegraphics[width=\columnwidth]{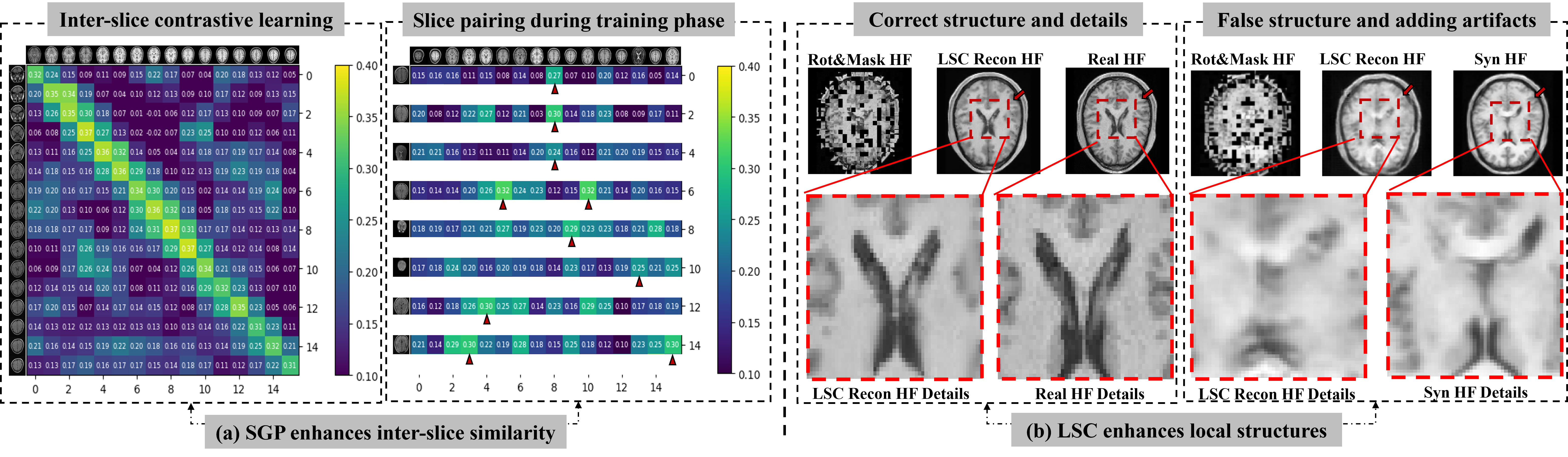}}
\caption{Visualization result of SGP and LSC process. (a) SGP enhances inter-slice similarity by pre-training on sequential low-field and high-field MRI data and matching the most similar slices within a randomly shuffled batch. (b)  LSC enhances local structures by recovering fine image details from locally masked and rotated images.}
\label{fig3}
\end{figure*}
\begin{figure*}[t]
\centerline{\includegraphics[width=\columnwidth]{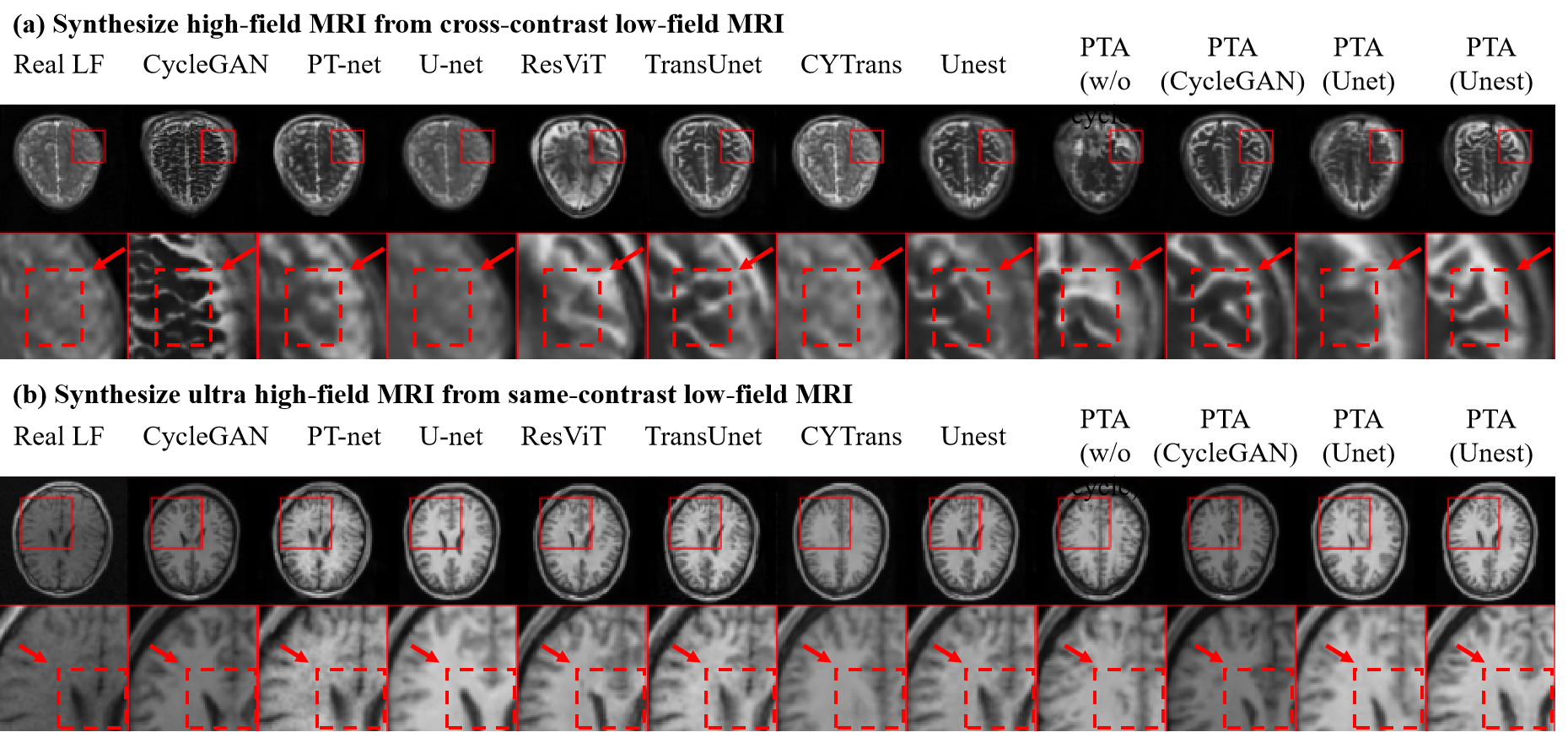}}
\caption{Visualization result of different baselines for exemplar regions (top and bottom row of each panel) (a) Synthesis of high-field MRI data from cross-contrast low-field MRI data. (b) Synthesis of ultra high-field MRI from same-contrast low-field MRI.}
\label{fig4}
\end{figure*}

\section{Method}
\textbf{Problem Formulation:} Let the low-field MRI data and high-field MRI data (magnitude images) be sampled from two different distributions, $X$ and $Y$, respectively. The goal is to train a generator $G_\theta$ that maps $X$ to a synthesized high-field MRI domain $Y'$, where $\theta$ represents the generator's parameters. However, this task is challenged by inter-slice misalignment and local detail discrepancies.
Additionally, a discriminator $D_\phi$ is defined to evaluate whether the synthesized image $Y'$ belongs to the high-field MRI domain $Y$, with $\phi$ as its parameters. The hypothesis space for this mapping is formulated as:
\begin{equation}
\mathcal{H}=\left\{\left(G_{\theta}, D_{\phi}\right) \mid G_{\theta}: \mathcal{X} \rightarrow \mathcal{Y}, \; D_{\phi}: \mathcal{Y} \rightarrow [0,1], \; \theta \in \Theta, \; \phi \in \Phi\right\}
\end{equation}
where $\Theta$ and $\Phi$ denote the parameter sets for the generator and discriminator, respectively. The overall objective is to determine the optimal parameters $\theta^{*}$ and $\phi^{*}$ by minimizing the combined loss function:
\begin{equation}
\theta^{*},\phi^{*}=\arg \min_{\theta \in \Theta,\phi \in \Phi} \mathcal{L}_{G}(G(X),Y) + \mathcal{L}_{D}(G(X),Y)
\end{equation}
where $\mathcal{L}_G$ and $\mathcal{L}_D$ represent the generator and discriminator losses, respectively. 

\textbf{Slice-wise Gap Perception:}
The slice-wise gap perception module aims to mitigate misalignment between low-field and high-field MRI slices by leveraging contrastive learning. It ensures spatial consistency by matching corresponding slices in unpaired MRI datasets. The contrastive loss function is defined as::
\begin{equation}
\mathcal{L}_{\text{SGP}} = -\frac{1}{N} \sum_{i=1}^N \log \frac{\scriptstyle
  \exp\left(\operatorname{sim}\left(f(S_{\mathrm{LF}}^{(i)}), f(S_{\mathrm{HF}}^{(i)})\right) / \tau \right)
}{\scriptstyle
  \sum_{j=1}^N \exp\left(\operatorname{sim}\left(f(S_{\mathrm{LF}}^{(i)}), f(S_{\mathrm{HF}}^{(j)})\right) / \tau \right)
}
\end{equation}
where $S_{LF}^{(i)}$ is the input hybrid image, $S_{HF}^{(i)}$ is the corresponding positive hybrid image (same pair type), $S_{HF}^{(j)}$ is a negative hybrid image (different pair type), $f(\cdot)$ represents the feature extractor, and $\tau$ is a temperature scaling factor.
This loss encourages the model to minimize the feature distance between matching hybrid images and maximize the distance between non-matching ones.\\
\begin{table}[t!]
\centering
\caption{Comparison of methods based on FID, IS, and MS-SSIM (Mix Contrast and Single Contrast). Bold indicates the best performance of our PTA configuration}
\label{tab:comparison}
\resizebox{\columnwidth}{!}{
\begin{tabular}{l|ccc|ccc}
\hline\hline
 \multicolumn{1}{c}{}& \multicolumn{3}{c}{\textbf{LISA, M4RAW $\rightarrow$ fastMRI, HCP1200}} & \multicolumn{3}{c}{\textbf{M4RAW $\rightarrow$ HCP1200}}\\
\hline
 \multicolumn{1}{c}{}&\multicolumn{3}{c}{\textbf{Mix-Contrast: (T1, T2)$\rightarrow$(T1, T2)}} & \multicolumn{3}{c}{\textbf{Single-Contrast: (T1)$\rightarrow$(T1)}}\\
\hline
\textbf{Methods}  & \multicolumn{1}{c}{\textbf{FID}$\downarrow$} & \multicolumn{1}{c}{\textbf{IS}$\uparrow$} & \multicolumn{1}{c|}{\textbf{MS-SSIM}$\downarrow$} & \multicolumn{1}{c}{\textbf{FID}$\downarrow$} & \multicolumn{1}{c}{\textbf{IS}$\uparrow$} & \multicolumn{1}{c}{\textbf{MS-SSIM}$\downarrow$} \\
\hline
\multicolumn{7}{l}{\textbf{Paired Methods}}\\\hline
Syn-GAN\cite{SynGAN}    &171.009  &1.131$\pm$0.196  & 0.989$\pm$0.026   & 156.058 &1.071$\pm$0.001  &0.995$\pm$0.003 \\
ESR-GAN\cite{esrgan}    &184.045  &1.627$\pm$0.064  &0.406$\pm$0.178 &165.725  &1.708$\pm$0.097  &0.597$\pm$0.200 \\\hline
\multicolumn{7}{l}{\textbf{Unpaired Methods}}\\\hline
ResViT\cite{resvit}  &  56.956& 1.943$\pm$0.077 & 0.269 $\pm$0.220   &61.487& 1.896$\pm$0.115 & 0.382 $\pm$0.139 \\
PTNet\cite{ptnet}  &81.096  &2.027$\pm$0.110  &0.180$\pm$0.158    &101.128  &1.921$\pm$0.089  &0.359$\pm$0.110  \\
CyTrans\cite{cytran}  &85.837  &1.486$\pm$0.075  & 0.191$\pm$0.149   &41.192  &1.868$\pm$0.082  & 0.437$\pm$0.201\\
TransUnet\cite{transunet}  &54.427
  &2.040 $\pm$0.067  &0.234$\pm$0.150    &90.996
  &1.713 $\pm$0.066  &0.339$\pm$0.120  \\
\hline
Cycle-GAN\cite{Cyclegan}   &  61.470& 2.068$\pm$0.207 &   0.201$\pm$0.140 &   43.861& 1.890$\pm$0.093 &   0.362$\pm$0.139\\
\textbf{Our PTA (Cycle-GAN\cite{Cyclegan}})     &\textbf{26.963}  &\textbf{2.140$\pm$0.119}  &0.223$\pm$0.189    &\textbf{16.892}  &1.840$\pm$0.094  &0.363$\pm$0.145 \\
\hline
U-net\cite{unet}  &  44.929& 1.928$\pm$0.105 &   0.232$\pm$0.193 & 33.770& 1.833$\pm$0.088 &   0.442$\pm$0.212\\
\textbf{Our PTA (U-Net\cite{unet}})     &40.156  &1.951$\pm$0.086  &0.205$\pm$0.195    &32.409  &1.916$\pm$0.170  &0.403$\pm$0.187 \\
\hline
Unest\cite{UNest}  &49.020  &1.755$\pm$0.091  &0.208$\pm$0.193    &51.748  &1.827$\pm$0.049  &0.372$\pm$0.138 \\
\textbf{Our PTA (Unest\cite{UNest}})    &48.838  & 1.853$\pm$0.080 &\textbf{0.198$\pm$0.168}    &47.979  &\textbf{1.933$\pm$0.098}  &\textbf{0.324$\pm$0.135} \\
\hline\hline
\end{tabular}}
\end{table}

\begin{figure*}[t]
\centerline{\includegraphics[width=\columnwidth]{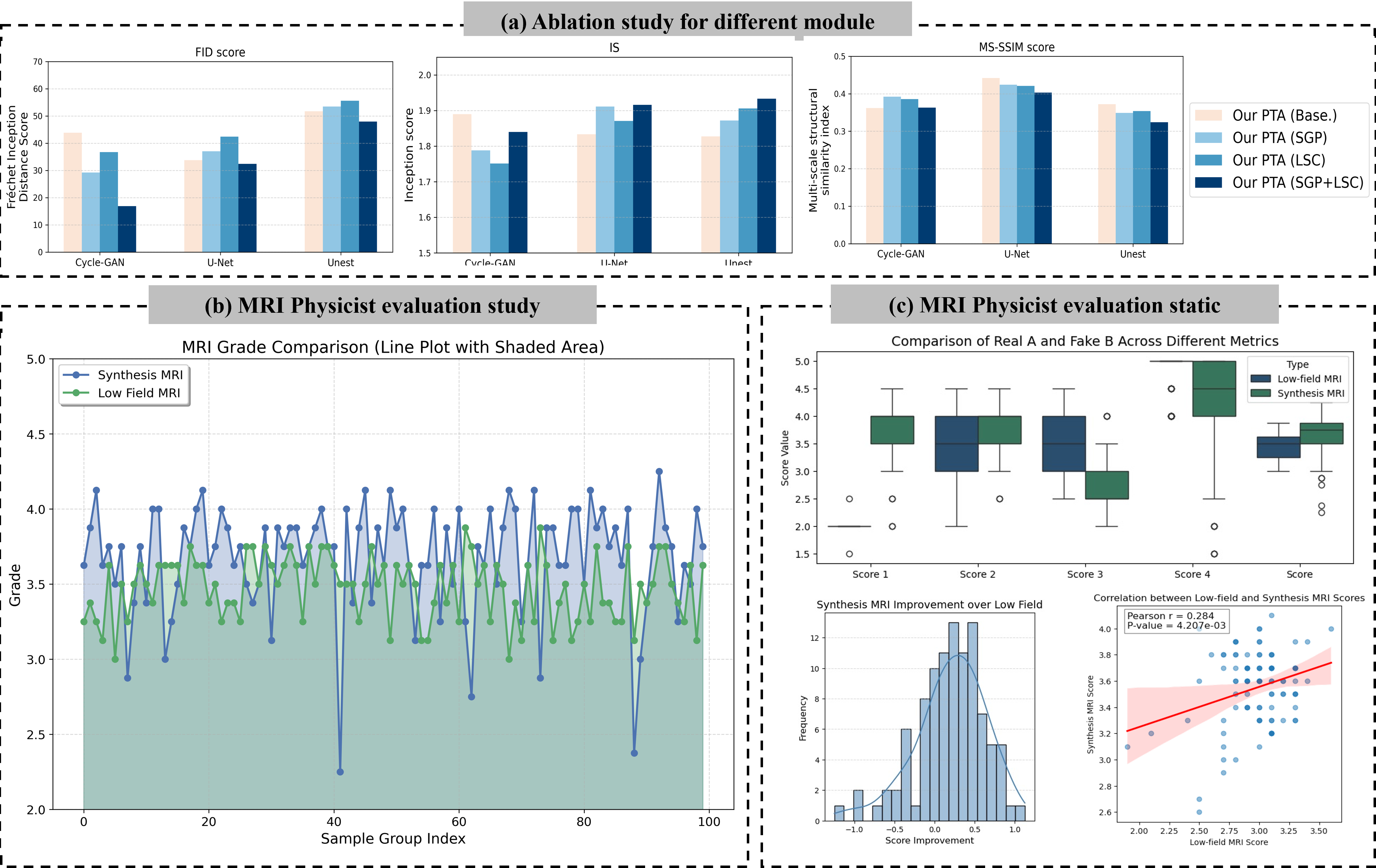}}
\caption{Ablation study and MRI physicist evaluation of our PTA.(a) Performance comparison of different PTA configurations through an ablation study. (b) MRI physicist evaluation study, comparing quality scores between low-field MRI and synthesized MRI using a 5-point Likert scale. (c) MRI physicist evaluation statistics, including box plots, histograms, and correlation analysis.}
\label{fig5}
\end{figure*}
\textbf{Local-Structure Correctness:}
To better capture fine-grained structural details in high-field MRI data, we adopt a pre-training strategy that enhances local feature learning. This method involves randomly rotating and masking local regions of high-field MRI data, followed by training a specific network to reconstruct the original structure. To disrupt local structures, the original high-field MRI data is divided into non-overlapping local blocks, which undergo the following transformations: Random Rotation $\psi_{rot}$: A subset of blocks is rotated by 90°, 180°, or 270°. Random Masking $\phi_{mask}$: Another subset of blocks is randomly masked, occluding structural details. The transformed image is then processed by a task-specific network, which aims to restore the original high-field MRI data by learning local feature relationships. The process is optimized by minimizing the reconstruction error, defined as:
\begin{equation}
\mathcal{L}_{\text {LSC}}=\left\|Y_{\text {syn}}-E_{T}\left(\phi_{mask}\psi_{rot} (Y_{\text {syn}}\right))\right\|_{2}^{2}
\end{equation}
where $Y_{syn}$ represents the synthesized high-field MRI data, and $E_T(\cdot)$ denotes the task-specific reconstruction network. By minimizing 
$\mathcal{L}_{LSC}$, the model enhances its ability to perceive local structural details, ensuring that high-field anatomical features are reconstructed in the most realistic way.\\
\textbf{Cycle Synthesis Adversarial Training}
In the training stage, we employ a SGP Network to reduce slice discrepancies between low-field and high-field MRI data. To ensure anatomical consistency, a cycle consistency loss $\mathcal{L}_{cycle}$ enforces low-field $\to$ high-field $\to$ low-field reconstruction. To further refine details, a LSC Network  enhances local anatomical structures in synthesized images. Finally, it employs an adversarial loss $\mathcal{L}_{adv}$ that promotes realism by distinguishing real and synthesized high-field MRI data.
\begin{equation}
\mathcal{L}_{all} = \lambda_{1}*\mathcal{L}_{\text{syn}} + \lambda_{2} *\mathcal{L}_{\text{cycle}} + \lambda_{3}*\mathcal{L}_{\text{adv}} ,
\end{equation}
where $\lambda_{1} = 0.5$, $\lambda_{2} = 0.2$, and $\lambda_{3} = 0.3$ are tuned to balance synthesis quality, structural consistency, and adversarial realism.


\section{Experiment}
\textbf{Dataset:} The datasets used in this study are categorized into low-field MRI datasets and high-field MRI datasets. The low-field datasets include M4RAW \cite{m4raw} and LISA dataset\cite{lisa}. The M4RAW dataset comprises brain MRI scans from 183 healthy volunteers, acquired using a 0.3T MRI scanner. The LISA dataset contains over 300 pediatric T2-weighted scans obtained with a 0.064T MR scanner. The high-field datasets include fastMRI \cite{fastMRI} and HCP1200 \cite{HCP}. The fastMRI dataset consists of 6,970 fully sampled brain MRIs, acquired on 1.5T and 3T scanners. The HCP1200 dataset provides 7T MRI data from 184 healthy participants, offering ultra-high-resolution brain imaging.\\
\textbf{Metrics:} The performance of PTA is evaluated using three key metrics: 1) Fréchet Inception Distance (FID) measures the distributional similarity between generated and real MRI data, where lower FID values indicate higher fidelity \cite{fid}. 2) Inception Score (IS) assesses both the quality and diversity of synthesized images, with higher IS values reflecting more realistic and varied outputs \cite{is}. 
 3) Multiscale Structural Similarity Index (MS-SSIM) evaluates perceptual similarity at multiple scales by analyzing structural, luminance, and contrast features across different resolutions \cite{ms-ssim}. Lower MS-SSIM values suggest better diversity in the generated images.\\
\textbf{Comparison Experiment:} 
Table \ref{tab:comparison} compares PTA framework with two paired and seven unpaired image synthesis methods, reports performance in FID, IS, and MS-SSIM. Despite the trade-off between FID and MS-SSIM, the PTA framework achieved the lowest FID (16.892) and the highest IS (1.933) at low MS-SSIM (no more than 0.363) in low-field to ultra-high-field tasks. This shows that the image quality is still excellent while maintaining diversity. The PTA framework achieves the lowest MS-SSIM (0.324) at a lower FID (49.979), further confirming that the PTA framework can generate diversity while ensuring fidelity. Figure \ref{fig4} shows that PTA framework generates images with finer textures and better structural consistency than baseline methods, preserving anatomical details and features (details pointed by red arrows). These results highlight its effectiveness in high-field MRI synthesis and potential for downstream tasks.\\
\textbf{Ablation Study:} 
Evaluation of PTA framework is conducted through an ablation study, testing the effects of SGP, LSC, and their combination (SGP $+$ LSC). As shown in Figure \ref{fig5} (a), incorporating these modules improves PTA framework performance across FID, IS, and MS-SSIM. SGP reduces slice misalignment, improving FID (43.861$\rightarrow$29.239) and IS (1.833$\rightarrow$1.911), ensuring better spatial consistency. LSC refines local structures, lowering MS-SSIM (0.372$\rightarrow$0.354) and enhancing fine-grained details. Their combination (SGP $+$ LSC) achieves the best trade-off, yielding the lowest MS-SSIM (0.324), highest IS (1.933), and competitive FID (47.979), balancing fidelity and diversity. \\
\textbf{MRI Phyisicist Evaluation:}
Since unpaired generation tasks are difficult to verify, we introduce an MRI physicist evaluation experiment to verify the synthesis quality. It is conducted following standard MRI brain image quality scoring guidelines. The evaluation criteria included Signal-to-Noise Ratio, Contrast-to-Noise Ratio, Spatial Resolution, and Global Artifacts, rated on a 5-point Likert scale ranging from Non-diagnostic (1) to Excellent (5). This assessment provides an objective measure of the quality of synthesized ultra high-field MRI data. Fig. \ref{fig5} (b) presents a line plot comparison of MRI physicist evaluation scores. The results indicate that most synthesized MRI data outperform their low-field counterparts, with higher scores across most samples. However, some synthesized images exhibit corrupted structures or artifacts, leading to a number of lower scores, though they remain within an acceptable range (3\%). Fig. \ref{fig5} (c) provides a statistical analysis of the evaluation through different visualizations. The box plot compares score distributions between low-field MRI and synthesized MRI, showing that synthesized images generally receive higher ratings but with some overlap. The histogram illustrates the frequency of score improvements (mean improvement = 0.201), confirming a shift toward higher-quality MRI synthesis from low-field data to high-field data. The correlation plot further demonstrates a positive relationship between low-field MRI quality and improvements in synthesized MRI (Pearson r = 0.284). This highlights that the PTA framework is based on the quality improvement of the original low-field images. This indirectly reflects the ability of the PTA framework to enhance image fidelity.

\section{Conclusion}


In this work, we proposed a Pretext Task Adversarial (PTA) learning framework to tackle the challenges of synthesizing realistic high-field MRI data from low-field inputs, and augmenting scarce datasets. Our framework integrates slice-wise gap bridging, structure correctness and adversarial training, enhancing anatomical accuracy in the generated images. Extensive experiments on low-field to ultra high-field task demonstrated the effectiveness of our approach (16.892 for FID, 1.933 for IS score and 0.324 for MS-SSIM scores). Despite its promising results, our work still has limitations. The application is currently limited to MRI data, restricting generalizability to other imaging modalities. Our future work will focus on addressing these limitations by improving the robustness to diverse inputs and extending it to other medical imaging modalities.

%
%
%
\bibliographystyle{splncs04}
\bibliography{Miccai/ref}
%




\end{document}